\documentclass{article}

\usepackage{graphicx}
\usepackage{multirow}
\usepackage{caption}
\usepackage{amssymb, amsmath}
\usepackage{url}
\usepackage{authblk}

\begin{document}
\title{Low-ordered Orthogonal Voxel Finite Element with INT8 Tensor Cores for GPU-based Explicit Elastic Wave Propagation Analysis}
\author[1]{Tsuyoshi Ichimura}
\author[1]{Kohei Fujita}
\author[2]{Muneo Hori}
\author[1]{Maddegedara Lalith}
\affil[1]{Earthquake Research Institute and Department of Civil Engineering, The University of Tokyo, Japan}
\affil[2]{Research Institute for Value-Added-Information Generation, Japan Agency for Marine-Earth Science and Technology, Japan}
\date{}
\maketitle
\begin{abstract}
Faster explicit elastic wavefield simulations are required for large and complex three-dimensional media using a structured finite element method. Such wavefield simulations are suitable for GPUs, which have exhibited improved computational performance in recent years, and the use of GPUs is expected to speed up such simulations. However, available computational performance on GPUs is typically not fully exploited, and the conventional method involves some numerical dispersion. Thus, in this paper, we propose an explicit structured-mesh wavefield simulation method that uses INT8 Tensor Cores and reduces numerical dispersion to speed up computation on GPUs. The proposed method was implemented for GPUs, and its performance was evaluated in a simulation experiment of a real-world problem. The results demonstrate that the proposed method is 17.0 times faster than the conventional method.
\end{abstract}

\section{Introduction}

Explicit simulation of elastic wavefields, which can be computed without solving matrix equations, is often used to evaluate the dynamic response of large and complex three-dimensional media through sequential analysis and to estimate internal structures through optimization via many simulations. Thus, further speedup is desired for conducting such large and many-case analyses. In this paper, we focus on increasing the speed of explicit elastic wavefield simulations with structured finite elements, which are suitable for generating numerical models. In the standard finite element method, the generation of the finite element model can be a bottleneck in the analysis process. However, finite element models can be generated automatically for the structured finite element method by using cubic elements that are finer than the elements used in standard finite element analysis and allowing some approximation of geometry, thereby making it suitable for analyzing large and complex media. Thus, it is used in seismic analysis \cite{ref_eq1,ref_eq2} and the estimation of the internal structure of structures \cite{ref_ultra1}. There is a strong need for a method that can be executed many times in a short period because detailed simulations and model optimizations are desired in such analyses.

Explicit wavefield simulations using structured finite element methods, where continuous memory access is dominant, are well-suited for processing on GPUs and are expected to be executed quickly on recent GPUs with improved computational performance. However, the lumped mass matrix approximation is applied to the low-order structured finite elements that are conventionally used in such analyses, and this results in numerical dispersion, which increases computational cost due to the need to use smaller element sizes. In addition, some GPUs are equipped with acceleration mechanisms, e.g., Tensor Cores \cite{TensorCore}, and their use in deep learning and other applications is progressing. However, the use of Tensor Cores in physics-based simulations is limited because their high-speed operations are predicated on transforming operations into a somewhat special form, and they are only used in some cases (e.g., to obtain a coarse solution in implicit simulations \cite{TCFEM1,ref_PASC2020}. See \cite{ref_MixedPrecisionSurvey} for a review of mixed-precision algorithms for linear numerical algebra in general). In particular, unlike the implicit solution method, the explicit solution method does not allow for processes that refine the coarse solution. Therefore, the development of such algorithms is challenging because the results must be equivalent to those of ordinary FP64 calculations while effectively using Tensor Cores. However, Tensor Cores have high computational performance; thus, further increases in processing speed can be expected if the hardware can be exploited effectively.

Therefore, this paper proposes an explicit structured finite element elastic wavefield simulation method that uses INT8 Tensor Cores, which is more accurate and faster than the conventional method. The remainder of this paper is organized as follows. In Section \ref{sct2}, we present an explicit structured finite element wavefield simulation method that solves the problems of conventional methods and is suitable for INT8 Tensor Core computation. Section \ref{sct3} describes the implementation of the proposed method on a GPU and presents the details of the performance of the proposed method using a realistic analysis model. We show that the proposed method can realize an analysis that yields results that are equivalent to those of the conventional method at higher speed. Finally, the paper is concluded in Section \ref{sct4}.

\section{Low-ordered Orthogonal Voxel Finite Element with INT8 Tensor Cores}
\label{sct2}

In this study, we target the governing equation of a linear dynamic elastic body:
\begin{equation}
\label{GE:ORG}
\rho \ddot{\mathbf{u}}-(\nabla \cdot \mathbf{c} \cdot \nabla) \cdot \mathbf{u}=\mathbf{f},
\end{equation}
where $\rho$, $\mathbf{u}$, $\mathbf{c}$, and $\mathbf{f}$ denote the density, displacement,  elasticity tensor, and body force, respectively. In addition, $(\dot{~})$ and $\nabla$ denote the temporal and spatial differential operators, respectively. In standard finite element analysis, the target domain is decomposed into small elements $e$, and the basis functions $\phi^{\alpha}$ are defined on these elements. By substituting $\mathbf{u}=\sum \mathbf{u}^{\alpha} \phi^{\alpha}$ into the functional in Eq.~(\ref{GE:ORG}) and taking the stationary condition, the following linear equation is obtained for the unknown coefficients $\mathbf{u}^{\alpha}$:
\begin{equation}
\label{DGE:ORG}
\mathbf{K} \mathbf{u}+\mathbf{M} \mathbf{\ddot{u}}=\mathbf{F}.
\end{equation}
Here, $\mathbf{u}$ is a vector assembled using $\mathbf{u}^\alpha$. $\mathbf{K}$ and $\mathbf{M}$ are a stiffness matrix and mass matrix assembled using the element stiffness matrix $\mathbf{K}_e$ and the element mass matrix $\mathbf{M}_e$ obtained for each element, respectively.

\begin{figure}[tb]
\centering
\includegraphics[width=0.8\hsize]{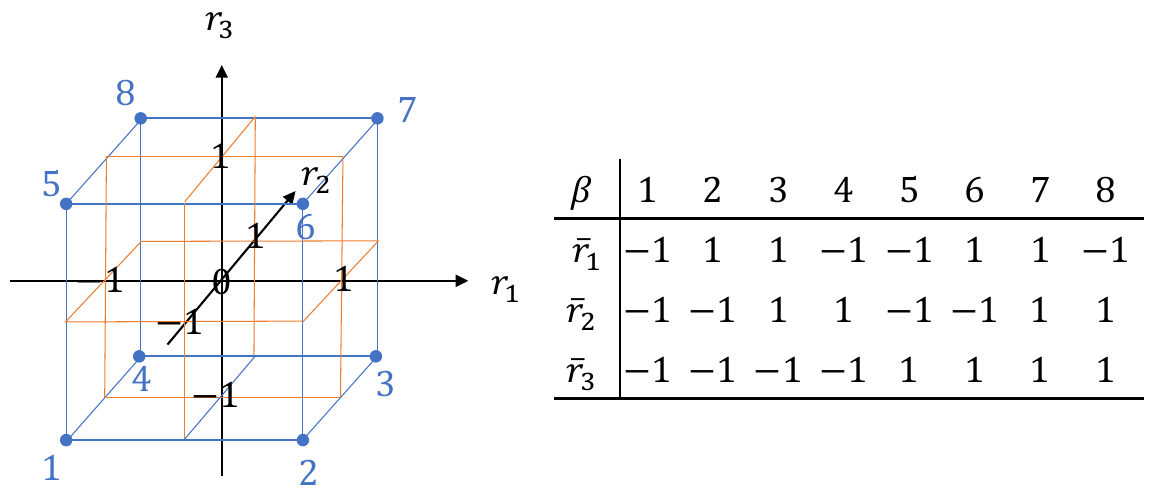}
\caption{Definition of a voxel element. For a cube of size $ds$ in each direction, a local coordinate $r_1r_2r_3$ is defined ($-1 \le r_1 \le 1$, $-1 \le r_2 \le 1$, $-1 \le r_3 \le 1$). The local node number ranging from 1--8 is defined on each node in the element. $\phi^{\beta}$ is the basis function for node $\beta$.}
\label{fig:defelem}
\end{figure}

In a typical structured finite element analysis, the domain of interest is divided into cubic subdomains of lengths $ds$ per side, and these are used as elements (hereafter referred to as voxel elements). As a result, a finite element model can be generated easily by simply covering a three-dimensional region with a structured grid of constant width $ds$ and evaluating the physical properties of each element. Thus, it is frequently used for the static and dynamic analysis of regions with complex geometry and property distributions or for geometry optimization. However, the object is modeled with cubes of uniform size; thus, it is necessary to reduce $ds$ to improve the accuracy of modeling geometry, which frequently results in a problem with large degrees of freedom. When using such voxel elements, low-order elements are frequently used, i.e., it is common to use basis functions
\begin{equation}
\phi^{\beta} = -\frac{1}{8} (r_1+\bar{r}_1) (r_2+\bar{r}_2) (r_3+\bar{r}_3)~~~(\beta=1,2,...,8)\nonumber
\end{equation}
for node $\beta$, which is often used for hexahedral elements (Figure~\ref{fig:defelem} shows the definition of the local coordinates, local node numbering, and the definition of $\bar{r}_1$, $\bar{r}_2$, $\bar{r}_3$). However, in this case, the element mass matrix
\begin{equation}
\mathbf{M}_e = \rho (\phi ^\beta \phi ^{\beta '})_e,\nonumber
\end{equation}
becomes nondiagonal, which necessitates solving a large-scale matrix equation to obtain $\mathbf{u}$ in Eq.~(\ref{DGE:ORG}). Note that $(~)_e$ indicates volume integration in each element. To solve this problem, an approximation that concentrates the off-diagonal terms to the diagonal terms (lumped mass matrix approximation) is frequently applied, which leads to a diagonal mass matrix $\mathbf{M}_e^{ap}$, where the off-diagonal terms are approximated as 0, and the diagonal terms are approximated as $\rho/8(1)_e$. Consequently, for example, if we approximate the acceleration using the central difference method with a time-step width of $dt$, Eq.~(\ref{DGE:ORG}) at time step $it$ becomes
\begin{equation}
\mathbf{K} \mathbf{u}^{it}+\mathbf{M}^{ap} (\mathbf{u}^{it+1}-2 \mathbf{u}^{it} +\mathbf{u}^{it-1})/dt^2=\mathbf{F}^{it},
\label{eqtimehistory}
\end{equation}
which enables explicit computation of $\mathbf{u}^{it+1}$. This lumped matrix approximation is widely used when solving dynamic problems because eliminating the need to solve large matrix equations reduces the analysis cost significantly. This analysis method is referred to as VFEM in this paper.

In VFEM, the lumped mass matrix approximation causes numerical dispersion, which is particularly noticeable when the distance of the propagation of waves increases. To suppress such numerical dispersion, a VFEM using an orthogonal basis function has been proposed for the voxel elements \cite{ref_OVFEM}. Using the orthogonal basis functions, the mass matrix becomes diagonal without any approximation to the voxel elements. We formulate OVFEM using the Hellinger-Reissner functional:
\begin{equation}
\label{HRfunctional}
J (\mathbf{u}, \mathbf{\sigma})=\int_V \int_T -\frac{1}{2} \rho \dot{\mathbf{u}} \cdot \dot{\mathbf{u}}-\frac{1}{2} \mathbf{\sigma}:\mathbf{c}^{-1}:\mathbf{\sigma} + \mathbf{\sigma}:(\nabla \otimes \mathbf{u}) - \mathbf{u}\cdot \mathbf{f}~ \mathrm{d}v \mathrm{d}t.
\end{equation}
Here, $\mathbf{\sigma}$ denotes stress, which can be expressed as $\mathbf{\sigma} = \sum \mathbf{\sigma}^{\beta} \psi^{\beta}$ using the basis function $\psi^{\beta}$ and unknown coefficients $\mathbf{\sigma}^{\beta}$. By taking the stationary condition of Eq.~(\ref{HRfunctional}) for $\mathbf{u}^{\alpha}$ and $\mathbf{\sigma}^{\beta}$, a discretized form of the governing equation can be obtained in the form of Eq.~(\ref{DGE:ORG}). Here, $\mathbf{K}$ and $\mathbf{M}$ become matrices assembled using the element stiffness matrix
\begin{equation}
\label{eq:OVFEMstiff}
\mathbf{K}_e^o = (\psi^{\beta} \nabla \phi^{\alpha})^t_e \cdot (\psi^{\beta} \psi^{\beta'})_e^{-1} \mathbf{c}\cdot (\psi^{\beta} \nabla \phi^{\alpha})_e,
\end{equation}
and the element mass matrix
\begin{equation}
\label{eq:OVFEMmass}
\mathbf{M}_e^o = \rho (\phi^{\beta} \phi^{\beta'})_e,
\end{equation}
respectively. While the same local coordinates $-1\le r_1 \le1$, $-1\le r_2 \le1$, $-1\le r_3 \le1$ and node placement of VFEM (Fig.~\ref{fig:defelem}) are used, the following Heaviside function $H$ is used in OVFEM:
\begin{equation}
\label{eq:OVFEMbasis1}
\phi^{\beta} = H(\bar{r}_1 r_1) H(\bar{r}_2 r_2) H(\bar{r}_3 r_3)~~~(\beta=1,2,...,8), \\
\end{equation}
\begin{equation}
\psi^{1}=1, 
\psi^{2}=r_1, 
\psi^{3}=r_2, 
\psi^{4}=r_3,
\psi^{5}=r_1 r_2, 
\psi^{6}=r_2 r_3, 
\psi^{7}=r_1 r_3.
\label{eq:OVFEMbasis2}
\end{equation}
As shown in the definition of Eq.~(\ref{eq:OVFEMbasis1}), the element mass matrix of Eq.~(\ref{eq:OVFEMmass}) becomes diagonal, and a stable solution is obtained because the discontinuous functions $H$ are used together with smooth polynomial functions. Thus, the use of $\mathbf{M}_e^o$ allows us to solve Eq.~(\ref{DGE:ORG}) explicitly without approximation. OVFEM is expected to be more accurate than VFEM due to the suppression of numerical dispersion by the absence of the lumped mass approximation. This indicates that OVFEM may enable a coarser discretization size compared to VFEM, which can reduce the analysis costs of three-dimensional dynamic problems considerably.

Some GPUs are equipped with acceleration mechanisms that can perform certain forms of matrix products at high speed. For example, the A100 80 GB PCIe GPU \cite{ref_A100} has Tensor Cores with peak performance of 624 TOPS for 8-bit integer matrix operations, which is significantly higher than the 9.7 TFLOPS peak performance of its FP64 cores. Therefore, effective use of Tensor Cores is expected to speed up analysis, and Tensor Cores are being used in deep learning and other applications. However, to apply Tensor Core operations efficiently, it is necessary to formulate an algorithm such that the amount of matrix operations of a particular form is dominant while guaranteeing the accuracy of the operations, which limits the applicability to only a few physics-based simulations. In particular, with the implicit method, Tensor Cores can be used as a substitute for part of the analysis by using Tensor Cores to obtain a coarse solution and then refining it to reduce the analysis cost; however, it is difficult to apply such a refining process to the explicit method. Thus, the part of the algorithm to be computed by Tensor Cores must demonstrate accuracy that is equivalent to that of FP64, and the difficulty in algorithm development lies in the fact that speedup must be achieved in consideration of the computing performance, data transfer, and other factors.

Therefore, we propose TCOVFEM, which is an OVFEM transformed to take advantage of the INT8 Tensor Core arithmetic performance with guaranteed arithmetic accuracy. Generally, solids can be treated as a linear isotropic dynamic elastic body in the most basic approximation; thus, the formulation is presented using a linear isotropic dynamic elastic body as an example. This method can be applied to the ultrasonic wave propagation analysis shown in the application examples presented in this paper and to the analysis of wavefields in other linear isotropic elastic bodies. The key is determining how to make $\mathbf{K} \mathbf{u}$, which has the highest analysis cost in Eq.~(\ref{DGE:ORG}), suitable for Tensor Core operations while guaranteeing FP64 accuracy. First, we consider the computation of $(\psi^{\beta} \nabla \phi^{\alpha})_e$ among the operations given in Eq.~(\ref{eq:OVFEMstiff}). Note that there is arbitrariness in the selection of $\psi^{\beta}$ in Eq.~(\ref{eq:OVFEMbasis2}); thus, we select a polynomial that follows the deformation mode of the cube, which leads to an expression within the range of low-precision numbers by proper normalization. In contrast, the computation of $(\psi^{\beta} \psi^{\beta'})_e^{-1} \mathbf{c}$ requires care. Generally, $\mathbf{c}$ does not fall within the range of low-precision numbers (even if it is linear isotropic); thus, it must be treated appropriately. In this section, we focus on the independence of the properties of linear isotropic elastic bodies and derive a formulation that falls within the range of 8-bit integers. For linear isotropic elastic bodies, $\mathbf{c}$ can generally be expressed by the product of the volume modulus $\kappa$, the shear modulus $G$, and a constant that fits into low-precision arithmetic. Here, by defining $\bar{\mathbf{D}}^{\kappa}= 3 \frac{\partial (\psi^{\beta} \psi^{\beta'})_e^{-1} \mathbf{c}}{\partial \kappa} ds^3$ and $\bar{\mathbf{D}}^{G}= 3 \frac{\partial (\psi^{\beta} \psi^{\beta'})_e^{-1} \mathbf{c}}{\partial G} ds^3$ as the contribution of $\kappa$ and $G$ to $(\psi^{\beta} \psi^{\beta'})_e^{-1} \mathbf{c}$, a constant $24 \times 48$ matrix
\begin{equation}
\mathbf{K}_e^\mathrm{INT8}=\begin{pmatrix}{\mathbf{K}_e^{\kappa}, \mathbf{\bar{K}}_{e}^{G}}\end{pmatrix}, \nonumber 
\end{equation}
where 
\begin{eqnarray}
\mathbf{K}_{e}^{\kappa}&=&256/3 \bar{\mathbf{B}}^t \bar{\mathbf{D}}^{\kappa} \bar{\mathbf{B}} ds^4, \nonumber \\
\bar{\mathbf{K}}_{e}^{G}&=&128 \bar{\mathbf{B}}^t \bar{\mathbf{D}}^{G} \bar{\mathbf{B}} ds^4 - 128 \mathbf{I}, \nonumber
\end{eqnarray}
can be constructed. Here, $\bar{\mathbf{B}}$ is an equivalent strain displacement matrix and $\mathbf{I}$ is a $24 \times 24$ identity matrix. Using this transformation, the core kernel, which has the highest analysis cost in finite element analyses, can be computed as follows:
\begin{equation}
\label{GE:TC}
\mathbf{K}^o_e \mathbf{u}_e=1/256 \kappa ds (\mathbf{K}_{e}^\mathrm{INT8} \bar{\mathbf{u}}_e + 256/3 G/\kappa \mathbf{u}_e).
\end{equation}
Here, $\bar{\mathbf{u}}_e= \begin{pmatrix} \mathbf{u}_e, 2/3 G/\kappa \mathbf{u}_e \end{pmatrix}$.

In Eq.~(\ref{GE:TC}), we convert most of the analysis cost into operations that fall within the 8-bit integer range; however, to extract the full performance of Tensor Cores, it is necessary to implement Tensor Core computations in consideration of both the operation speed and the data transfer cost. Here, we present a concrete implementation of the fast Tensor Core-based computation of the dense matrix-vector product $\mathbf{K}_{e}^{\mathrm{INT}8} \bar{\mathbf{u}}_e$. Here, $\mathbf{K}_{e}^{\mathrm{INT}8}$ is a 24$\times$48 constant matrix with values in the 8-bit integer range (-128 -- 127), and $\bar{\mathbf{u}}_e$ is a vector with a length of 48 in FP64. Note that conversion between floating-point numbers and integers incurs a large cost; thus, we consider fast computation algorithms that minimize the number of variable conversions while reducing the data transfer cost of the computation.

First, to save the number of significant digits during integer arithmetic expansion, the element right-hand side vector is scaled as follows:
\begin{equation}
\bar{\mathbf{u}}_{es} = \frac{1}{s_e} \bar{\mathbf{u}}_e.
\label{intscaling}
\end{equation}
Here, $s_e = \max_{i=1,2,...,48} | \bar{u}_{ei} |$, where $\bar{u}_{ei}$ is the $i$-th component of $\bar{\mathbf{u}}_e$. This scales each component of $\bar{\mathbf{u}}_e$ to be within $\pm 1$. Then, $\bar{\mathbf{u}}_{es}$ is expanded with the product of FP64 constants times integer value vectors:
\begin{equation}
\bar{\mathbf{u}}_{es} = \sum_{i=1}^N \frac{1}{a^i}  \bar{\mathbf{u}}_{es\mathrm{INT}(i)},
\label{intcomp1}
\end{equation}
where $N$ is the number of integer expansion stages, and $\bar{\mathbf{u}}_{es\mathrm{INT}(i)}$ is a vector comprising 48 integers:
\begin{eqnarray}
\bar{\mathbf{u}}_{es\mathrm{INT}(1)} &\underset{\mathrm{INT}}{\Leftarrow}& a \bar{\mathbf{u}}_{es}, \label{intcomp12} \\
\bar{\mathbf{u}}_{es\mathrm{INT}(2)} &\underset{\mathrm{INT}}{\Leftarrow}& a \left\{ a \bar{\mathbf{u}}_{es} -   \bar{\mathbf{u}}_{es\mathrm{INT}(1)} \right\}, \\
\bar{\mathbf{u}}_{es\mathrm{INT}(3)} &\underset{\mathrm{INT}}{\Leftarrow}& a \left\{ a^2 \bar{\mathbf{u}}_{es} - a \bar{\mathbf{u}}_{es\mathrm{INT}(1)} - \bar{\mathbf{u}}_{es\mathrm{INT}(2)}  \right\}. \label{intcomp2}\\
&...& \nonumber 
\end{eqnarray}
Here, $\underset{\mathrm{INT}}{\Leftarrow}$ denotes truncation to integer values. Using these integer vectors, the element matrix-vector product is computed as follows:
\begin{equation}
\mathbf{K}_{e}^{\mathrm{INT}8} \bar{\mathbf{u}}_e =
s_e \sum_{i=1}^N \frac{1}{a^i} \mathbf{K}_e^{\mathrm{INT}8} \bar{\mathbf{u}}_{es\mathrm{INT}(i)}.
\label{intcomp3}
\end{equation}

\begin{figure}[t]
\begin{center}
\begin{small}
\includegraphics[width=\hsize]{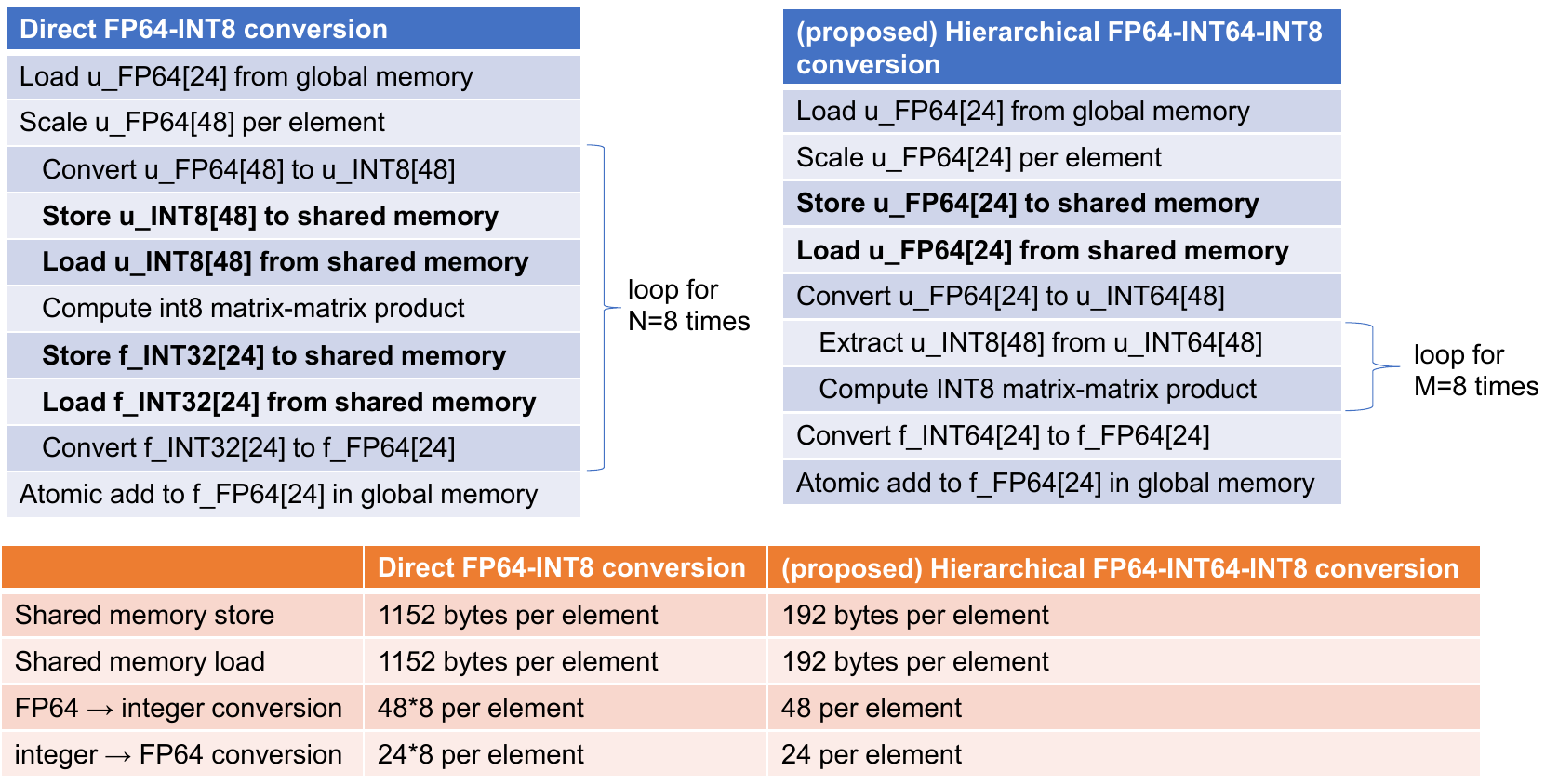}\\
\caption{Computation of matrix-vector products using INT8 Tensor Cores with direct FP64-INT8 conversion and the proposed hierarchical FP64-INT64-INT8 conversion. Here, 32 elements are computed using 32 threads per thread block.}
\label{figmatrixvectoralgorithm}
\end{small}
\end{center}
\end{figure}

The above Eqs.~(\ref{intcomp1})--(\ref{intcomp3}) can be expanded and calculated with integers with an arbitrary number of bits. For example, using $a=2^7$, all integer operations can be performed with 8-bit integers. However, in this case, Eqs.~(\ref{intcomp12})--(\ref{intcomp2}) must be calculated for the $N$ stages, which requires many conversions between floating-point variables and integers, and, therefore, reduces the computational performance. For example, $N=8$ stages are required to realize 56-bit accuracy which is equivalent to FP64's 52 fraction bits, which requires many conversion operations. Thus, in this study, the number of conversions between the FP64 variables and integers is reduced using 64-bit integers to perform the conversion only once and by using a hierarchical expansion in which the 64-bit integer calculation is further expanded using 8-bit integer operations. In particular, Eqs.~(\ref{intcomp1})--(\ref{intcomp12}) are expanded using 64-bit integers with $a=2^{56}$ and $N=1$, and the main computation $\mathbf{K}_e^{\mathrm{INT}8} \bar{\mathbf{u}}_{es\mathrm{INT}(1)}$ (denoted $\mathbf{K}_e^{\mathrm{INT}8} \bar{\mathbf{u}}_{es\mathrm{INT}64(1)}$) is expanded using $M=8$-stages of computations using 8-bit integers. Here, using $b=2^7$, $\bar{\mathbf{u}}_{es\mathrm{INT}64(1)}$ is expanded as
\begin{equation}
\bar{\mathbf{u}}_{es\mathrm{INT}64(1)} =  \sum_{j=1}^M b^{j-1} \bar{\mathbf{u}}_{es\mathrm{INT}8(1,j)} 
\label{int8comp1}
\end{equation}
using 8-bit integers, and $\mathbf{K}_e^{\mathrm{INT}8} \bar{\mathbf{u}}_{es\mathrm{INT}64(1)}$ is computed as
\begin{equation}
\sum_{j=1}^M b^{j-1} \mathbf{K}_e^{\mathrm{INT}8} \bar{\mathbf{u}}_{es\mathrm{INT}8(1,j)}.
\label{int8comp2}
\end{equation}
Figure~\ref{figmatrixvectoralgorithm} shows the computation flow of the matrix-vector products when directly converting data between FP64 and INT8 variables and when hierarchical FP64-INT64-INT8 conversion is used. While the data conversion and computation are looped $N$ times in the direct method, the proposed hierarchical method performs the data conversion outside of the $M$ loop. While the total computation accuracy is the same when $N=M$, the number of conversions between the integer and FP64 variables can be reduced by $N$-fold when using the hierarchical method.

\begin{figure}[t]
\centering
\includegraphics[width=0.8\hsize]{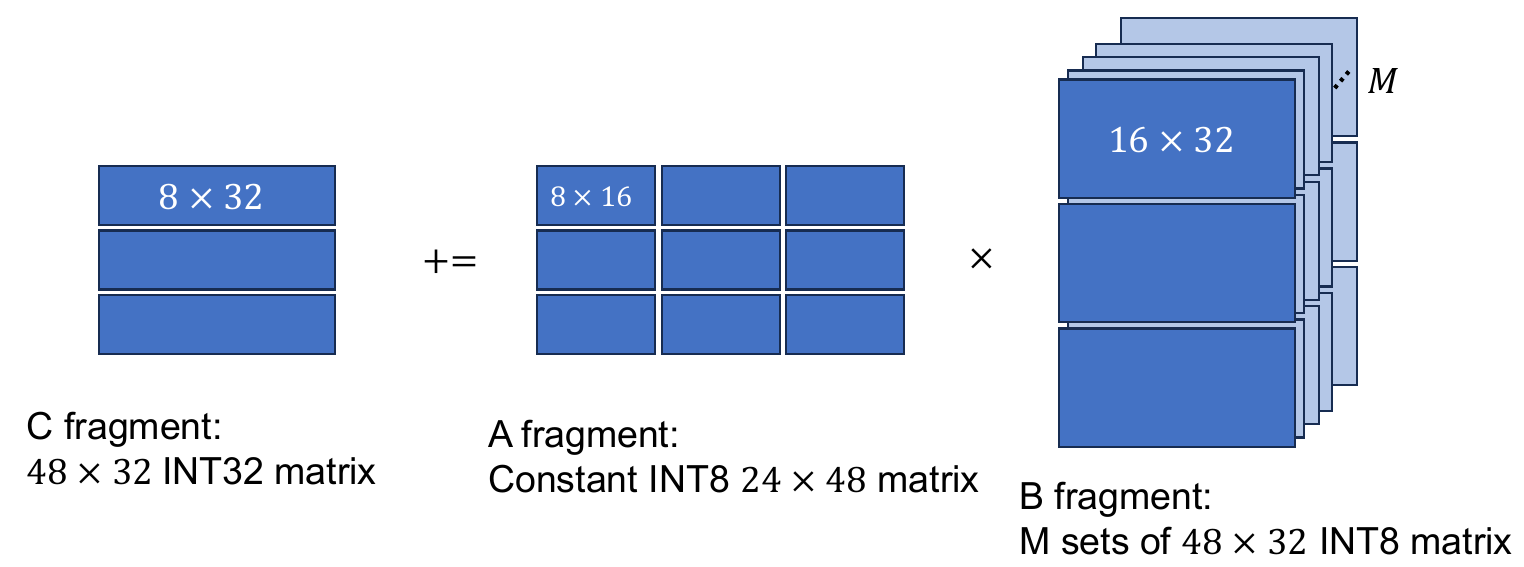}\\
\caption{Decomposition of (24$\times$48)$\times$(48$\times$32) matrix-matrix product into nine (8$\times$16)$\times$(16$\times$32) matrix-matrix products. Here, the A fragment can be reused throughout the $M$ sets of computations. Note that the results in the 32-bit integer C fragment are flushed every $M=3$ stages to an INT64 buffer to avoid overflow.}
\label{figmatrixmatrix}
\end{figure}

\begin{figure}[t]
\centering
\includegraphics[width=\hsize]{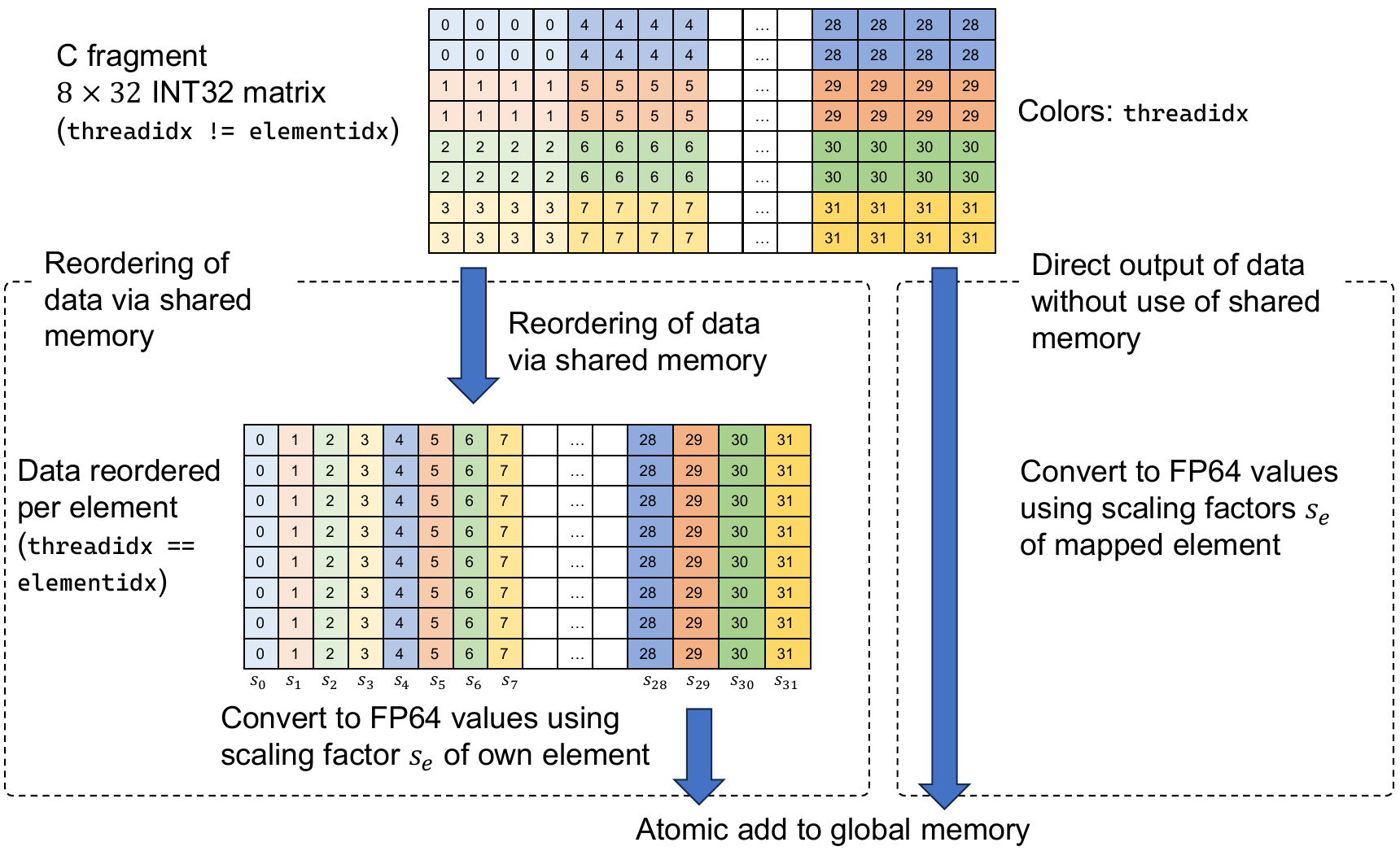}\\
\caption{Elimination of shared memory loads/stores by direct addition of results to global memory. Note that one out of the three C fragments is shown. 32 elements are computed using 32 threads per thread block.}
\label{figsharedmemory}
\end{figure}

Tensor Core operations are very fast when applied to matrix-matrix products of specific sizes; thus, we must transform the computation pattern of the matrix-vector products to matrix-matrix products suited for Tensor Cores. In addition, Tensor Core operations are very fast; thus, the cost of other operations must be suppressed to reduce time costs substantially. In particular, the data transfer from global memory and the data transfer between shared memory becomes a bottleneck. In the following, we describe how we map the matrix-vector computation to Tensor Cores and how we circumvent the data transfer bottlenecks.
\begin{itemize}
\item{Transformation of matrix-vector products to matrix-matrix products:}
While $\mathbf{K}_e^{\mathrm{INT}8} \bar{\mathbf{u}}_{es\mathrm{INT}8(1,j)}$ is a matrix-vector product of size (24$\times$48)$\times$48, we can compute 32 elements in a single thread block as a matrix-matrix product of size (24$\times$48)$\times$(48$\times$32). We compute this using INT8 Tensor Cores with $(n,m,k)=(8,32,16)$: C($8\times32$: INT32)=A($8\times16$: INT8)$\times$B($16\times32$: INT8). As shown in Fig.~\ref{figmatrixmatrix}, a (24$\times$48)$\times$(48$\times$32) matrix-matrix multiplication is decomposed into $3\times3=9$ parts, and each part is computed using $(8\times16)\times(16\times32)$ matrix-matrix multiplication. Here, the matrix $\mathbf{K}_e^{\mathrm{INT}8}$ is constant in the $j$-loop; thus, we can load the A fragment once and keep it in registers.
\item{Elimination of shared memory loads/stores by direct addition of results to global memory:}
As the register mapping of memory fragments for the matrices in Tensor Core operations is complex, an API is provided to exchange the input and output data between threads via shared memory. However, when this technique is used for the small-scale matrix-matrix multiplications targeted in this study, it frequently results in shared memory bottlenecks, which makes it difficult to fully utilize the available performance of the Tensor Cores. Thus, following the method presented in the literature \cite{ref_PASC2020}, we skip the remapping process using shared memory and directly output the data to global memory. As shown in Fig.~\ref{figsharedmemory}, rather than remapping the outputs from the Tensor Core operations and having each thread add element-wise results to the global memory, the results are added directly to the designated components in the global memory. Here, the scaling coefficient $s_e$ for each element is reflected in this procedure by sharing $s_e$ among the threads. Similarly, we can load the B fragments directly from the registers after conversion from INT64 value inputs by proper element mapping.
\item{Improving cache reuse when accessing global memory variables:}
The matrix-vector product calculation requires reading the nodal data $\mathbf{u}_e$ from global memory and adding the nodal result $\mathbf{f}_e$ to the global vector $\mathbf{f}$. In this study, 32 elements belonging to a single thread block are arranged in the $x$-direction to facilitate sequential cache access in the $x$-direction, and then each thread block performs calculations continuously in the $z$-direction. This allows the reuse of nodal data ($\mathbf{u}$ and $\mathbf{f}$) in cache, which results in an approximately fourfold reduction in the volume of global memory accesses compared to the case when elements are ordered randomly.
\end{itemize}

By employing a hierarchical method that reduces the number of conversions between the floating-point and integer variables, as well as using the data access reduction methods described above, we can expect high increases in speed compared to the standard FP64 computations without reduction in computational accuracy.

\section{Numerical Experiment}
\label{sct3}

\begin{figure}[t]
\centering
\includegraphics[width=\hsize]{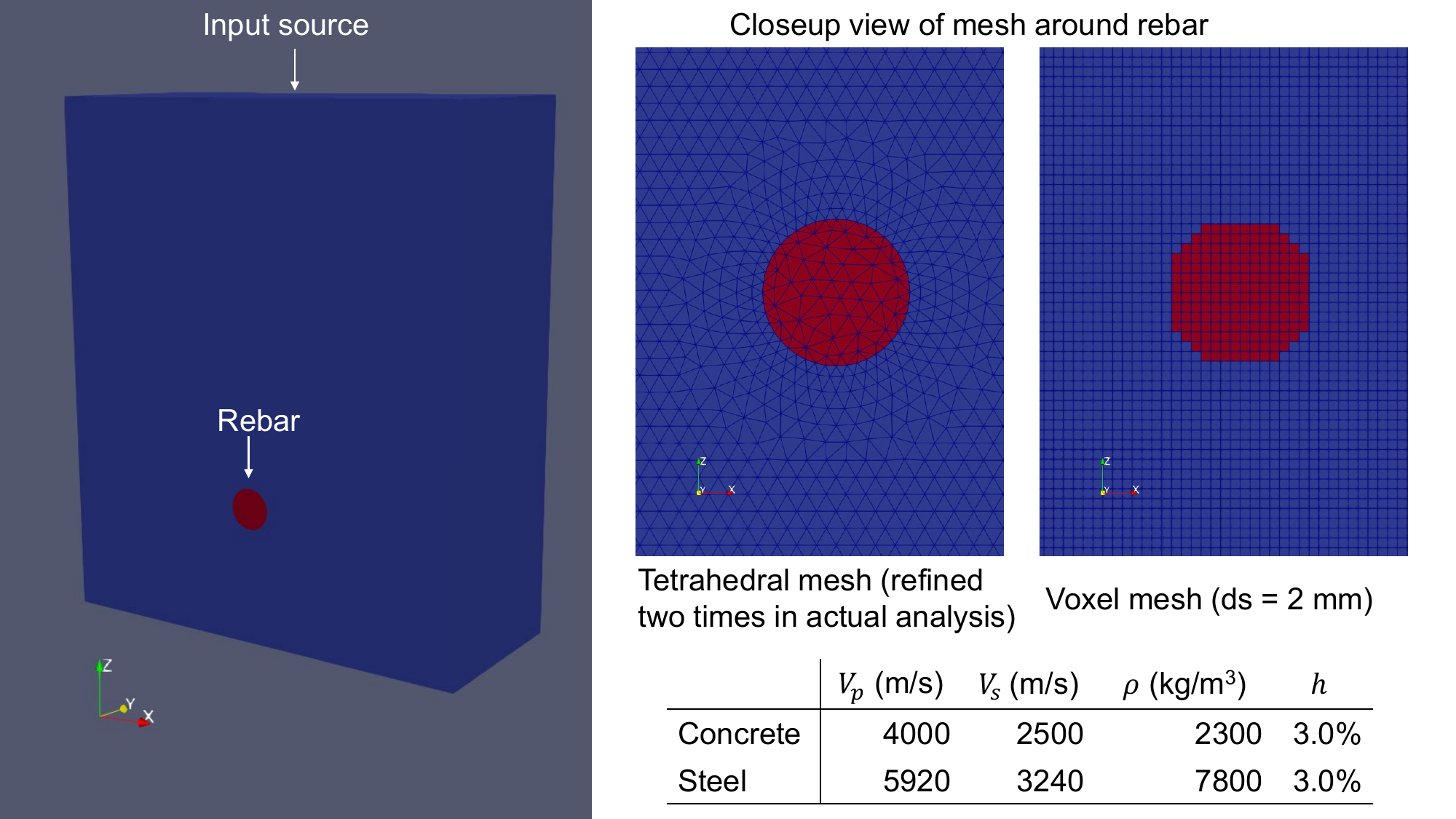}\\
\caption{Model used for the numerical experiment.}
\label{figmodel}
\end{figure}

In the following, the efficacy of the proposed TCOVFEM is demonstrated through ultrasonic analysis using the model illustrated in Fig.~\ref{figmodel}, which is based on the literature in \cite{ref_ultra_model}. The model is a rebar with a radius of 15 mm embedded in concrete. An impulse force with a center time of $4.096 \times 10^{-4}$ s and a center frequency of 112.5 kHz with a bandpass of 100--125 kHz is applied in the $z$-direction from the input point, and its response is calculated and stored for $8.192 \times 10^{-4}$ s at the observation points shown in Table~\ref{pointcoor}. The size of the model is $324 \times 128 \times 384$ mm and assumes a cylindrical rebar of radius 15 mm centered at $x=160$, $z=100$ mm that is penetrating the model in the $y$-direction. The four corner points at the bottom ($z=0$) of the model are fixed in three directions ($x$, $y$, and $z$). Rayleigh damping (100--125 kHz) is used to compute attenuation in the simulation. The following time history simulations were performed on a single A100 80 GB PCIe GPU \cite{ref_A100}, and except for the INT8 Tensor Core operation in the proposed TCOVFEM, all other operations were computed and stored using FP64 variables.

\begin{table}[t]
\caption{Coordinates of the input source and observation points.}
\label{pointcoor}
\centering
\begin{tabular}{c|ccc}
   & $x$ (mm) & $y$ (mm)&  $z$ (mm) \\  \hline
Source point & 156 & 72 & 384 \\
Obs. point 1 & 26 & 60 & 384 \\
Obs. point 2 & 60 & 60 & 384 \\
Obs. point 3 & 108 & 60 & 384 \\
Obs. point 4 & 144 & 60 & 384 \\
Obs. point 5 & 180 & 60 & 384 \\
Obs. point 6 & 216 & 60 & 384 \\
Obs. point 7 & 264 & 60 & 384 \\
Obs. point 8 & 300 & 60 & 384 \\ \hline
\end{tabular}
\end{table}

\begin{table}[t]
\caption{Comparison of relative error (TCOVFEM: proposed method; VFEM: explicit standard voxel FEM).}
\label{errorcomparison}
\centering
\begin{tabular}{c|rrr}
 & $ds$ (mm) & $dt$ (s) & $Err$ \\  \hline
TCOVFEM & 2.000 & $5.0\times10^{-8}$ & \bf{0.03947} \\
VFEM & 2.000 & $5.0\times10^{-8}$ & 0.28057 \\
VFEM & 1.500 & $2.5\times10^{-8}$ & 0.09579 \\
VFEM & 1.333 & $2.5\times10^{-8}$ & 0.05743 \\
VFEM & 1.200 & $2.5\times10^{-8}$ & \bf{0.03602} \\
VFEM & 1.091 & $2.5\times10^{-8}$ & 0.02343 \\
VFEM & 1.000 & $2.5\times10^{-8}$ & 0.01570 \\
VFEM & 0.500 & $1.25\times10^{-8}$ & 0.00104 \\ \hline
\end{tabular}
\end{table}

\begin{figure}[t]
\centering
\includegraphics[width=\hsize]{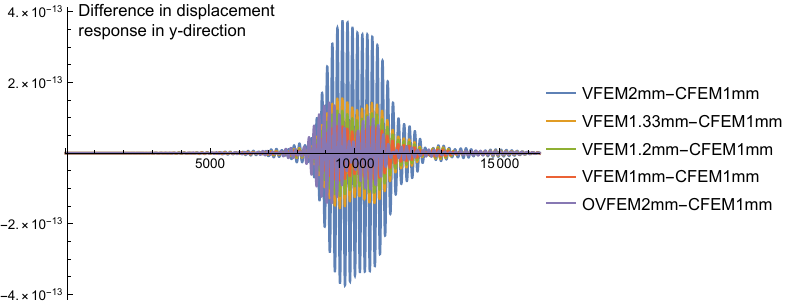}\\
\includegraphics[width=\hsize]{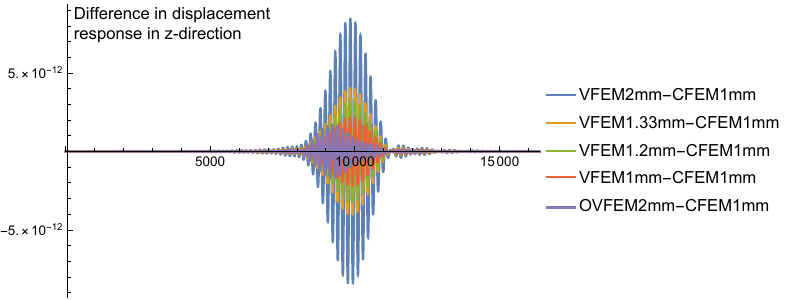}\\
\caption{Differences in time history displacement response at observation point 1 when compared with the reference solution ($y$- and $z$-directions with $dt=5\times10^{-8}$ s).}
\label{figtimehistory}
\end{figure}

First, a reference solution was generated using an implicit finite element analysis with second-order tetrahedral elements without the lumped mass matrix approximation (hereafter referred to as CFEM), and we compared the waveforms at the observation points to confirm the accuracy of the proposed TCOVFEM. As shown in Fig.~\ref{figmodel}, the finite element mesh used for the reference solution based on CFEM is sufficiently fine. For example, the area near the rebar is discretized with an element size of approximately 1 mm (i.e., the nodal spacing becomes approximately 0.5 mm as second-order tetrahedral elements are used). We used a model with 2-mm voxel elements for the proposed TCOVFEM, and to compare the accuracy with the conventional VFEM method, we conducted VFEM analysis for cases with seven element sizes (2.0, 1.5, 1.33, 1.2, 1.09, 1.0, and 0.5 mm). Table~\ref{errorcomparison} shows the error from the reference solution at all observation points, which is defined as follows for each analysis case.
\begin{equation}
Err = \frac{1}{n_c} \sum_{i=1}^{n_c} \frac{\sum_{j=1}^{n_t} (u_{i,j}^\mathrm{obs} - u_{i,j}^\mathrm{ref})^2}{\sum_{j=1}^{n_t} (u_{i,j}^\mathrm{ref})^2 }
\end{equation}
Here, $n_c$ denotes the number of observation channels (eight observation points $\times (x,y,z)$-components $=24$ channels), $n_t$ denotes the number of time steps, and $u_{i,j}^\mathrm{obs}$ and $u_{i,j}^\mathrm{ref}$ denote the computed and reference waves at time step $j$ for channel $i$, respectively. The results demonstrate that the proposed TCOVFEM can compute results with high accuracy compared to the conventional method. The error of CFEM using a coarser mesh (the length of an element edge is twice that of the mesh used in the reference solution) was 0.00022, which confirms that the reference solution computed using CFEM with the fine mesh is converged sufficiently. A comparison of the proposed TCOVFEM with the conventional VFEM indicates that the proposed method is more accurate than the VFEM in the case with the same discretization width. Here, the proposed TCOVFEM with $ds=2$ mm achieved accuracy that is equivalent to the VFEM with $ds=1.2$ mm. Figure~\ref{figtimehistory}, which visualizes the response in the $y$- and $z$-directions at observation point 1, also confirms that the proposed TCOVFEM with $ds=2$ mm is considerably more accurate than VFEM at $ds=2$ mm (even when viewed at individual observation points).

Next, we review the effect of INT8 Tensor Core operations on calculation accuracy. Here, we compare the accuracy when computing Eq.~(\ref{GE:TC}) without Tensor Cores using FP128, FP64, and FP32 operations, and computing Eq.~(\ref{GE:TC}) with INT8 Tensor Cores. Specifically, the calculations for multiplying the global stiffness matrix of the $ds=2$ mm cubic element model by random vectors with double precision values were computed with FP128, FP64, and FP32 operations, and these calculations were compared with the results computed using INT8 Tensor Core operations. Note that FP128 has 112 fraction bits, FP64 has 52 fraction bits, and FP32 has 23 fraction bits. In contrast, as the resolution of an 8-bit integer is 7 bits, the INT8 Tensor Core calculation is equivalent to 28 fraction bits for $N=4$ stages and 56 fraction bits for $M=8$ stages. Thus, it is expected that accuracy equivalent to that of FP32 can be obtained with $M=4$ stages and higher than that of FP64 with $M=8$ stages in the INT8 calculation. The results shown in Table~\ref{tableaccuracy} demonstrate that the accuracy of the INT8 operation increases with the number of stages, and that the INT8 operations with $M=8$ stages are equivalent to or higher than that of FP64. In addition, the INT8 operations with $M=4$ stages are equivalent to that of FP32. Thus, we use $M=8$ stages in the proposed TCOVFEM throughout the rest of this paper.

\begin{table}[t]
\centering
\caption{Comparison of computed values for one component of $\mathbf{K} \mathbf{u}$ computed for a random vector $\mathbf{u}$ on OVFEM with $ds=2$ mm. Values in bold letters are identical to the FP128 computation results.}
\label{tableaccuracy}
\begin{tabular}{c|cl}
Computation  type & Fraction bits & Value  \\  \hline
FP128 & 112 & {\bf 507813.690592559616827867910192902549} \\
FP64 & 52 & {\bf507813.6905925}632 \\
FP32 & 23 & {\bf50781}4.750 \\
INT8 ($M=4$) & 28 & {\bf507813}.7802133318 \\
INT8 ($M=8$) & 56 & {\bf507813.690592559}5  \\
\hline
\end{tabular}
\end{table}

Finally, we compare the elapsed time for the entire time series analysis and the matrix-vector product kernel (Table~\ref {tableperformance}). In the following, we refer to $\mathbf{K} \mathbf{u}^{it}$ in
\begin{equation}
\mathbf{u}^{it+1} \Leftarrow 2 \mathbf{u}^{it} - \mathbf{u}^{it-1} + dt^2 
\mathbf{M}^{-1} (\mathbf{F}^{it} + \mathbf{K} \mathbf{u}^{it}),
\end{equation}
which is obtained by transforming Eq.~(\ref{eqtimehistory}), as the matrix-vector product kernel, and the rest as other computations. We first compare the proposed TCOVFEM with $M=8$ stages (TCOVFEM INT8 ($M=8$)) with OVFEM without Tensor Cores (OVFEM FP64), which computes Eq.~(\ref{GE:TC}) as is with FP64 variables on the $ds=2$ mm model. In the model with $ds=2$ mm, the other computations took 4.62 s. By excluding this, we can evaluate the improved calculation time of the matrix-vector product operation when using the Tensor Cores. The matrix-vector product calculation was 43.3/9.62 = 4.5-fold faster using the Tensor Cores, which corresponds to 64.4 TOPS in computational performance. In contrast, the OVFEM matrix-vector product calculation using FP32 (without Tensor Cores), which had lower accuracy (Table~\ref{tableaccuracy}), took 15.5 s. Thus, the proposed TCOVFEM, which demonstrates computational accuracy that is equivalent to that of FP64 computations, is 1.6 times faster than the FP32 implementation. This improved calculation time for the matrix-vector product resulted in a 48.3/14.2 = 3.39-fold speedup of the overall time history simulation compared to the FP64-based OVFEM without Tensor Cores. We also compared the proposed method with the conventional VFEM. Note that VFEM uses the element-by-element (EBE) method in the explicit time integration code, which calculates the matrix-vector product in an element-by-element manner without keeping the entire stiffness matrix in memory (all calculations are performed in FP64). Here, the VFEM EBE (8-point integration) implementation in the paper \cite{ref_SCALA2020}, which achieves high computational performance as a regular EBE-based VFEM, was used for comparison. As shown in Table~\ref{errorcomparison}, elements with $ds=1.2$ mm are required for the VFEM to obtain accuracy that is equivalent to that of the proposed TCOVFEM with $ds=2$ mm sized elements. Consequently, since the number of elements is $(2.0/1.2)^3$ = 4.63 times greater, and the number of time steps is twice as large, the amount of operations increases such that the analysis takes 242.8 s even using this efficient VFEM implementation  (Table~\ref{tableperformance}). Thus, the proposed TCOVFEM is 242.8/14.2 = 17.0-fold faster than the conventional VFEM with equivalent accuracy in the wave calculation.

\begin{table}[t]
\centering
\caption{Comparison of elapsed time for wave propagation. Here, $dt=5\times10^{-8}$~s and 16,384 time steps were used for cases with $ds=2$ mm, and $dt=2.5\times10^{-8}$~s and 32,768 time steps were used for cases with $ds=1.2$ mm.}
\label{tableperformance}
\begin{tabular}{c|c|cc}
\multirow{2}{*}{Computation type} & $ds$  & \multicolumn{2}{c}{Elapsed time (s)} \\
& (mm) & Time-step loop & Matrix-vector product  \\  \hline
OVFEM FP64 & 2.0 & 48.3 & 43.3 [2.1 TFLOPS] \\
OVFEM FP32 & 2.0 & 18.9 & 15.5 [5.9 TFLOPS] \\
TCOVFEM INT8 ($M=8$) & 2.0 & 14.2 & 9.62 [64.4 TOPS] \\
\hline
VFEM FP64 & 1.2 &  242.8 & - \\
\hline
\end{tabular}
\end{table}

\section{Concluding Remarks}
\label{sct4}

We presented an explicit wavefield simulation method that uses structured finite elements with high speed and low numerical dispersion using INT8 Tensor Cores to effectively exploit the performance benefits of modern GPUs. A detailed comparison of the performance of the proposed and conventional methods on a real-world problem demonstrates that the proposed method utilizes GPU performance efficiently and that the proposed TCOVFEM (INT8 ($M=8$)) is 17.0-fold faster than the conventional VFEM. Tensor Cores, which are an acceleration mechanism in modern GPUs, are used in deep learning and other applications due to their high computing performance. However, the application of Tensor Cores in physics-based simulations remains limited because it must be brought into a specific format and data transfer must be performed without losing computing performance. In addition, the use of Tensor Cores to obtain a coarse solution and then refine it, as in the implicit method, is difficult in the explicit method, thereby making it even more challenging. In this context, the formulation and implementation presented in this paper, which exploits the performance advantages of Tensor Cores in explicit physics-based simulations, are general and expected to provide insights into accelerating other applications. The conversion of floating-point matrix-vector computations to small integer-based matrix-matrix computations is expected to be accelerated in other GPU architectures with acceleration mechanisms for integer-based matrix-matrix multiplication. Note that the purpose of this study was to demonstrate the basic performance of the proposed method; thus, the analysis was performed on only a single GPU. However, we expect that the proposed method can easily be extended to solve large-scale problems using multiple GPUs.

\end{document}